\documentclass[fleqn,usenatbib]{mnras}

\usepackage[T1]{fontenc}
\usepackage{ae,aecompl}

\usepackage{graphicx}	
\usepackage{amsmath}	
\usepackage{amssymb}	

\usepackage{newtxtext,newtxmath}

\usepackage[usenames]{xcolor}
\usepackage[normalem]{ulem}

\def\msun{{\rm M}_\odot}

\title[Central kinematics of M15]{MUSE narrow field mode observations of the central kinematics of M15}

\author[Usher et al.]{Christopher~Usher,$^{1}$\thanks{email: chris.usher@astro.su.se}
Sebastian Kamann,$^{2}$
Mark Gieles,$^{3,4}$
Vincent H\'enault-Brunet,$^{5}$ \newauthor
Emanuele Dalessandro,$^{6}$
Eduardo Balbinot,$^{7}$
Antonio Sollima$^{6}$
\\
$^{1}$Department of Astronomy, Oskar Klein Centre, Stockholm University, AlbaNova University Centre, SE-106 91 Stockholm, Sweden \\
$^{2}$Astrophysics Research Institute, Liverpool John Moores University, 146 Brownlow Hill, Liverpool L3 5RF, UK \\
$^{3}$ICREA, Pg. Llu\'{i}s Companys 23, E08010 Barcelona, Spain\\
$^4$Institut de Ci\`{e}ncies del Cosmos (ICCUB), Universitat de Barcelona (IEEC-UB), Mart\'{i} Franqu\`{e}s 1, E08028 Barcelona, Spain \\
$^{5}$Department of Astronomy and Physics, Saint Mary's University, 923 Robie Street, Halifax, NS B3H 3C3, Canada\\
$^{6}$INAF - Astrophysics and Space Science Observatory Bologna, Via Gobetti 93/3, I-40129 Bologna, Italy\\
$^{7}$Kapteyn Astronomical Institute, University of Groningen, Postbus 800, NL-9700AV Groningen, the Netherlands
}

\date{Accepted XXX. Received YYY; in original form ZZZ}

\pubyear{2019}

\begin{document}
\label{firstpage}
\pagerange{\pageref{firstpage}--\pageref{lastpage}}
\maketitle

\begin{abstract}
We present observations of the stellar kinematics of the centre of the core collapsed globular cluster M15 obtained with the MUSE integral field spectrograph on the VLT operating in narrow field mode.
Thanks to the use of adaptive optics, we obtain a spatial resolution of 0.1 arcsec and are able to reliably measure the radial velocities of 864 stars within 8 arcsec of the centre of M15 thus providing the largest sample of radial velocities ever obtained for the innermost regions of this system. 
Combined with previous observations of M15 using MUSE in wide field mode and literature data, we find that the central kinematics of M15 are complex with the rotation axis of the core of M15 offset from the rotation axis of the bulk of the cluster.
While this complexity has been suggested by previous work, we confirm it at higher significance and in more detail.
\end{abstract}




\section{Introduction}
\label{sec:intro}

The high stellar densities of globular clusters (GCs) have long proved a challenge in studying the kinematics of their resolved stars, particularly in their centres.
However, recent improvements in instrumentation and analysis techniques have allowed the kinematics of large numbers of GC stars to be studied.
Imaging from space using the Hubble Space Telescope has allowed proper motion measurements of GCs in their centres \citep[e.g.][]{2014ApJ...797..115B, 2015ApJ...803...29W} while {\it Gaia} provides high quality astrometry in cluster outskirts \citep[e.g.][]{2018A&A...616A..12G, 2018MNRAS.481.2125B, 2019MNRAS.485.1460S,2019MNRAS.489..623V}.
Recent advances in adaptive optics also provide the potential for useful astrometry in the centres of GCs \citep[e.g.][]{2016ApJ...833..111D, 2016A&A...595L...2M}.

Multiobject spectroscopy has allowed large samples of radial velocities to be measured  \citep[e.g.][]{2011A&A...530A..31L, 2015AJ....149...53K, 2018ApJ...860...50F} but conventional fibre or slit spectroscopy is limited to the lower density outskirts.
Integral field spectroscopy \citep[e.g.][]{2013ApJ...769..107L, 2014A&A...566A..58K, 2020MNRAS.492.3859D} allows spectra to be observed even in dense fields, especially when used with techniques to reliably extract spectra from blended stars \citep{2013A&A...549A..71K}.
With wide-field integral field spectrographs, large numbers of spectra can be obtained at once, with \citet{2018MNRAS.473.5591K} measuring the radial velocities of over 200 000 stars in 22 GCs using MUSE on the VLT \citep{2010SPIE.7735E..08B}.

M15 (NGC 7078) is an old ($\sim 13$ Gyr, e.g. \citealt{2010ApJ...708..698D, 2013ApJ...775..134V}), metal poor ([Fe/H] $= -2.33$ \citealt{2009A&A...508..695C}) GC with a mass of $\sim 5 \times 10^{5}\,\msun$ \citep[e.g.][]{2020MNRAS.492.3859D}.
According to the 2010 edition of the \citet{1996AJ....112.1487H} catalgoue, M15 lies 10.4 kpc from the Sun as well as 10.4 kpc from the Galactic centre.
Its steep, cuspy surface brightness profile strongly suggests that it is a core collapse GC \citep[e.g.][]{1986ApJ...305L..61D}.

To date, many groups investigated the kinematics of this GC through radial velocity \citep{1989ApJ...347..251P,1994AJ....107.2067G,2000AJ....119.1268G,2002AJ....124.3270G} and proper motion \citep{2014ApJ...797..115B} surveys.

Interest on the central kinematics of M15 \citep[e.g.][]{1976ApJ...208L..55N, 2002AJ....124.3270G, 2003ApJ...595..187M, 2003ApJ...582L..21B, 2011ApJ...732...67M, 2014MNRAS.438..487D} initially focused on the possible presence of an intermediate mass black hole (IMBH) in the cluster.
However, it is now generally considered unlikely that a core collapsed GC could host an IMBH as the excess kinetic energy that stars gain in interactions with the IMBH prevent their accumulation in the cluster centre \citep[e.g.][]{1977ApJ...218L.109I, 2003ApJ...582L..21B,2018MNRAS.473.4832G}.

More recent work has focused on the surprising detection of rotation in the central regions of M15.
\citet{2006ApJ...641..852V} found evidence for a fast spinning, kinematically decoupled core in the inner 4 arcsec of M15.
This result, already suggested by the analysis of \citet{1997AJ....113.1026G}, was later supported by the work of \citet{2013ApJ...772...67B} and \citet{2018MNRAS.473.5591K} who found the core of M15 shows a higher rotation and a different rotation axis than what is found at large radii.
This discovery was surprising given that ordered rotation should be quickly erased by the short relaxation time scale ($< 100$ Myr \citealt{1996AJ....112.1487H}) in the centre of M15.
\citet{2018MNRAS.473.5591K} found a clear connection between the angular momentum of a GC and its relaxation time with GCs with shorter relaxation times having lower specific angular momenta \citep[see also][]{2018MNRAS.481.2125B,2019MNRAS.485.4906D}.

In this paper we combine the techniques of adaptive optics and integral field spectroscopy, by analysing MUSE observations in the AO-supported narrow field mode to study the central kinematics of M15.
In Section \ref{sec:observations} we describe our observations and data reduction as well as our radial velocity measurements and the literature datasets we include in our analysis.
In Section \ref{sec:kinematics} we present the spatially resolved inner kinematics of M15 before discussing these kinematics in \ref{sec:discussion}.

\section{Observations and data reduction}
\label{sec:observations}
\subsection{MUSE narrow field mode}
We observed the centre of M15 using MUSE on the VLT in narrow field mode (NFM) as part of the NFM science verification (program 60.A-9492(A)) on the nights of 2018 September 13 and 14.
We observed four pointings in a 2 by 2 mosaic centred on the cluster centre.
Two of the pointings were observed for $5 \times 250$ s; the remaining two $4 \times 250$ s.
The DIMM seeing ranged from 0.5 to 1.2 arcsec.
The observations were all carried out using the GALACSI adaptive optics module \citep{2012SPIE.8447E..0JA, 2012SPIE.8447E..37S} which provides MUSE a 7.5 by 7.5 arcsec field-of-view with a spatial sampling of 0.025 arcsec in NFM.
In nominal mode, MUSE provides wavelength coverage between 4800 \AA{} and 9300 \AA{} with a gap between 5780 and 6050 \AA{} due to the notch filter used to suppress the light from the laser guide stars. 
MUSE has a near constant spectral FWHM of $\sim 2.5$ \AA{} across the wavelength range (corresponding to a resolution of $R \sim 1700$ at the blue end of the wavelength range and $R \sim 3500$ at the red end).
The datacubes were reduced using the standard MUSE pipeline \citep{2020A&A...641A..28W} in the same manner as in \citet{2018MNRAS.480.1689K}.

\begin{figure*}
\includegraphics[width=504pt]{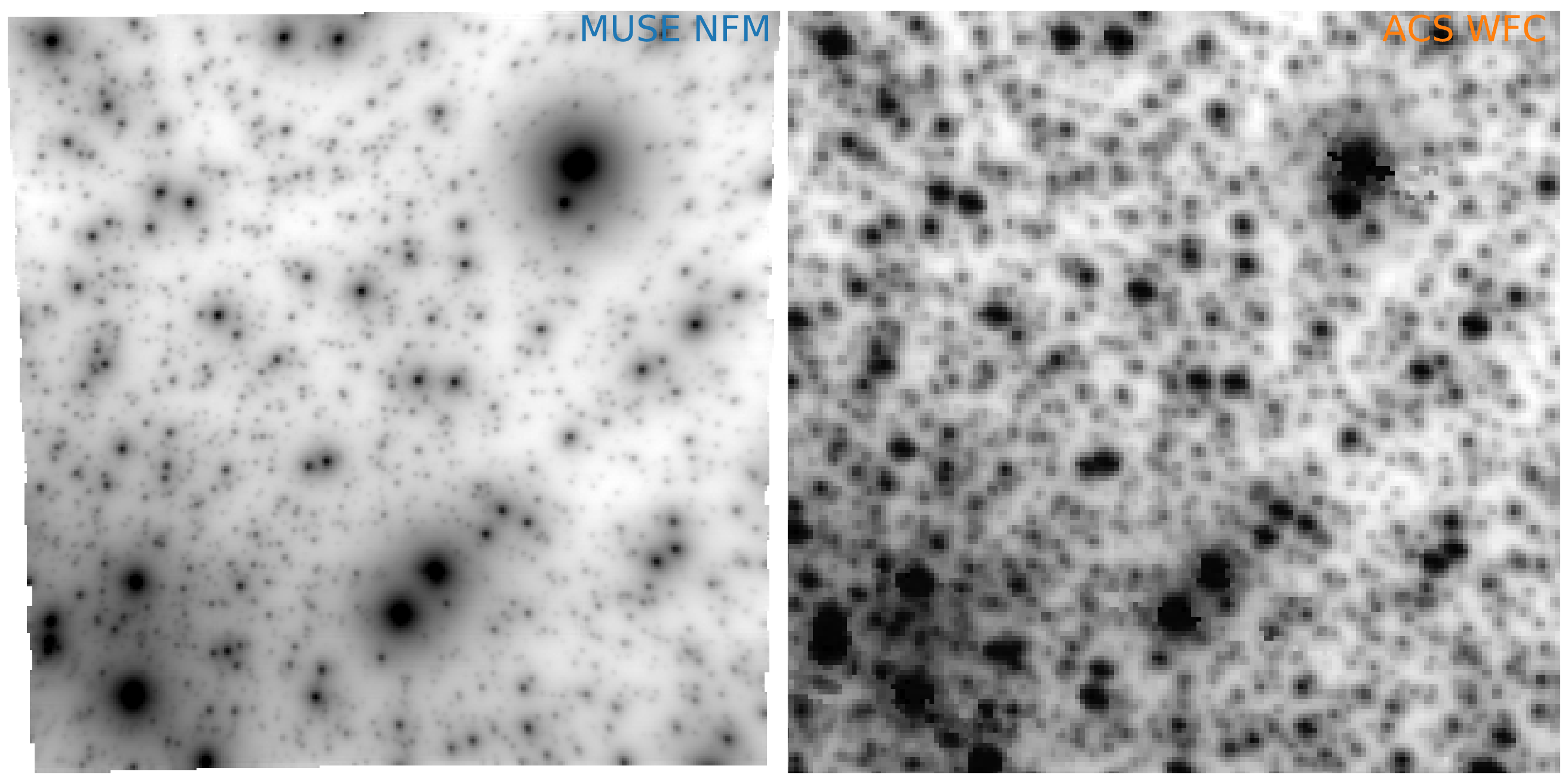}
\caption{Comparison of an $i$-band image created from one of the MUSE NFM pointings (left) and a HST ACS F814W image (right).
The MUSE NFM data shows a similar spatial resolution to the HST data.
Both images measure 7.98 by 7.88 arcsec.}
\label{fig:M15_iband}
\end{figure*}

We used \textsc{PampelMuse} \citep{2013A&A...549A..71K} to extract the stellar spectra from the MUSE datacubes.
\textsc{PampelMuse} determines the point spread function (PSF) from bright, isolated stars in the datacube and uses the PSF to extract spectra of all the sources in its input catalogue.
We utilised the \citet{2019ApJ...876...87B} HST ACS/HRC photometric catalogue as the input catalogue. During the extraction, we modelled the PSF using the combination of an outer Moffat and a central Gaussian profile. The idea behind this hybrid PSF model is that AO observations typically result in a PSF consisting of a diffraction-limited core surrounded by a seeing-limited halo. For the Gaussian component, we obtain a FWHM of $0\farcs08$ at $H_{\rm \alpha}$ and $0\farcs06$ at the calcium triplet.
We compare a $i$-band image created from one of the MUSE NFM pointings with the HST ACS/WFC F814W image from the ACS Globular Cluster Treasury Survey \citep[GO-10775][]{2007AJ....133.1658S} in Figure \ref{fig:M15_iband}.
In this wavelength region, the spatial resolution of the MUSE and HST data is is similar, with the MUSE PSF showing a sharper core but broader wings than the HST PSF.

Initial radial velocities were obtained by cross correlating the extracted spectra with synthetic templates from the PHOENIX library of \citet{2013A&A...553A...6H}. In order to find a suitable template for each spectrum, we compared the photometry of \citet{2019ApJ...876...87B} to an isochrone from the MIST \citep{2016ApJ...823..102C} database, adopting a cluster metallicity of $[{\rm Fe/H}]=-2.37$ \citep{1996AJ....112.1487H}. The comparison provided us with estimates of the effective temperature and surface gravity of each star with a MUSE spectrum available, which were in turn used to select a synthetic template for the cross correlation.

In the final step of the spectrum analysis, we processed each spectrum with \textsc{Spexxy} \citep{2016A&A...588A.148H}. The code performs a full spectrum fit against the same library of synthetic spectra used for the cross correlation to measure the radial velocity, effective temperature, and metallicity of each star. Initial guesses for the fit were obtained from the isochrone comparison described above as well as from the cross correlation.

As detailed in \citet{2016A&A...588A.149K}, the radial velocities obtained with \textsc{Spexxy} tend to be more precise than those measured via cross correlation, hence we gave preference to the former.
We obtained reliable radial velocities for 864 stars.
This is a significant increase compared to previous studies in the number of stars with radial velocities in the central few arcsec of M15 (see below).
Owing to the large range in brightness of our target stars, spectra were extracted over a wide range of signal-to-noise ratios, leading to very different uncertainties of the individual radial velocities. Our median accuracy is $7.3\,{\rm km\,s^{-1}}$, while the 16th and 84th percentiles of the uncertainties distribution are $2.6\,{\rm km\,s^{-1}}$ and $11.9\,{\rm km\,s^{-1}}$, respectively.

\subsection{Literature radial velocities}
We used the radial velocity catalogue from \citet{2018MNRAS.473.5591K} which used MUSE in wide field mode (WFM) to observe four pointings covering M15 out to $\sim 1$ arcmin at the same spectral resolution of the NFM dataset.
The MUSE WFM dataset was reduced and analysed in the same manner as the MUSE NFM data, providing radial velocities for 10420 stars.

We also utilised the literature radial velocities from \citet{2000AJ....119.1268G}, \citet{2002AJ....124.3270G} and \citet{2018MNRAS.478.1520B}.
\citet{2000AJ....119.1268G} used an imaging Fabry-Perot spectrograph together with the CFHT Adaptive Optics Bonnette \citep{1998PASP..110..152R} to obtain radial velocities of 104 stars within 20 arcsec of the centre of M15.
We note that these observations required 6.5 hours of exposure time.
\citet{2000AJ....119.1268G} combine this sample with their previous natural seeing Fabry-Perot observations and literature data to provide a dataset of 1779 radial velocities.
We further note that the astrometry of this catalogue is poor, especially in the outskirts of M15.

\citet{2002AJ....124.3270G} used multiple long slit HST STIS observations to mimic an integral field spectrograph.
They used 25 orbits of HST time to measure radial velocities for 64 stars.

The \citet{2018MNRAS.478.1520B} sample combines 273 radial velocities from APOGEE \citep{2018ApJS..235...42A} with their own radial velocity measurements of archival VLT FLAMES \citep{2002Msngr.110....1P} and Keck HIRES \citep{1994SPIE.2198..362V} spectra.
As these spectra were observed using traditional long slit or fibre spectroscopy, they all lie more than 30 arcsec away from the cluster centre.

\begin{figure*}
\includegraphics[width=504pt]{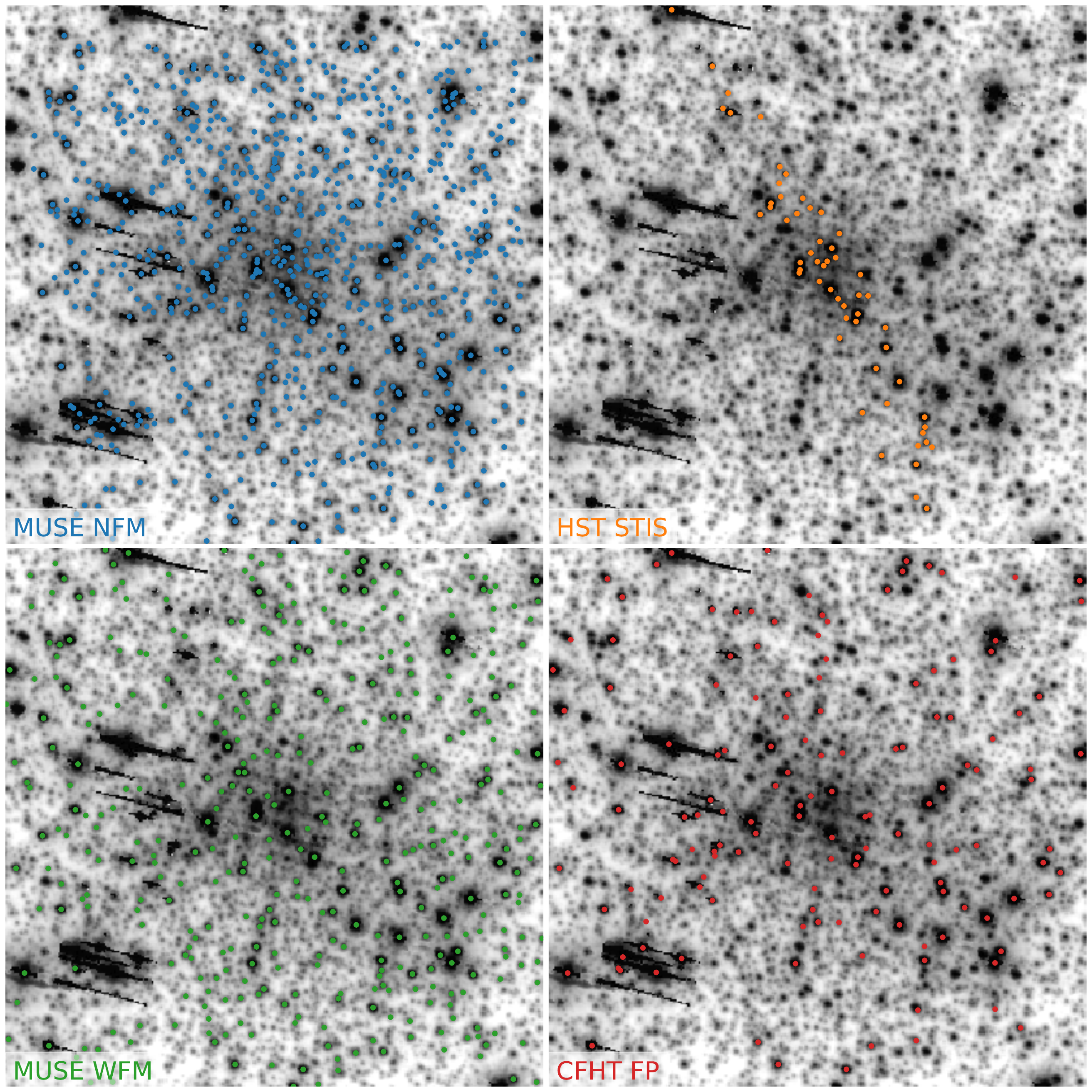}
\caption{Spatial comparison of the stars from different radial velocity sources.
Each panel shows the HST ACS F814W image of the central 15 by 15 arcsec of M15 over plotted by stars with our MUSE NFM measurements (top left in blue), the HST STIS measurements from \citet[][top right in orange]{2002AJ....124.3270G}, the MUSE WFM measurements from \citet[][bottom left in green]{2018MNRAS.473.5591K} and the CFHT Fabry-Perot measurements from \citet[][bottom right in red]{2000AJ....119.1268G}.
None of the velocity measurements from \citet{2018MNRAS.478.1520B} are within the inner 15 by 15 arcsec.
Our MUSE NFM observations allow us to measure radial velocities for significantly more stars compared to the MUSE WFM or CFHT FP observations in the same footprint and to measure radial velocities over a much larger area than the HST STIS observations.}
\label{fig:M15_data_sources}
\end{figure*}

\begin{figure}
	\includegraphics[width=240pt]{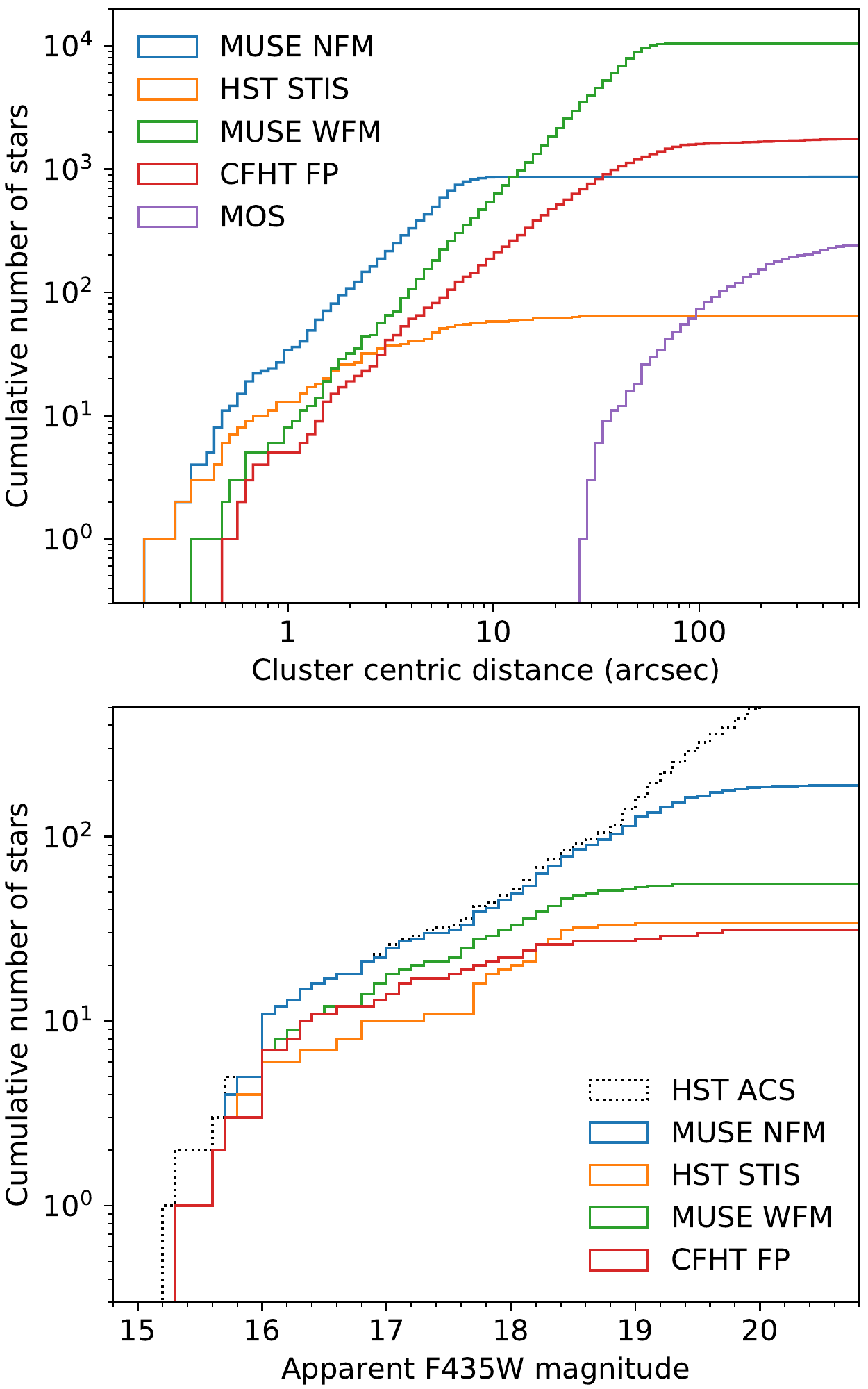}
	\caption{Top: cumulative radial distribution of stars from different radial velocity sources.
	Our MUSE NFM observations provide a dramatic increase in the number of stars with radial velocity measurements within the inner few arcsec of M15.
	Bottom: cumulative HST ACS F435W apparent magnitude distribution of stars from different radial velocity sources within 3 arcsec of the cluster centre.
	The black dotted line shows the magnitude distribution of all stars detected within 3 arcsec in the F435W imaging.
	The MUSE NFM observations allow us to measure radial velocities for fainter stars in the centre of M15 than previous studies.
	The MUSE NFM observations appear to be  as complete as the HST photometry to F435W magnitude of $\sim 19$.}
	\label{fig:cumulative}
\end{figure}

We compare the spatial distribution of stars from the first four catalogs in the inner 15 by 15 arcsec in Figure \ref{fig:M15_data_sources}.
We plot the radial distributions of stars from the five radial velocity catalogs in the upper panel of Figure \ref{fig:cumulative}.
Our MUSE NFM observations provide significantly more radial velocities within 8 arcsec compared to previous works.
The MUSE NFM observations provide 33 stars with radial velocities within 1 arcsec of the centre, 191 within 3 arcsec and 801 within 8 arcsec compared to 7, 57 and 363 stars respectively in the MUSE WFM observation within the same radii and 14, 57 and 173 stars within the same radii from the combined \citet{2000AJ....119.1268G} and \citet{2002AJ....124.3270G} samples.

In the lower panel of Figure \ref{fig:cumulative} we show the HST ACS/HRC F435W luminosity functions (using the photometry from \citealt{2019ApJ...876...87B}) for each of the radial velocity catalogs within 3 arcsec of the centre of M15.
Our MUSE NFM observations provide significantly more radial velocities for fainter stars than previous work.
We are able to measure radial velocities for virtually all stars in the ACS/HRC input catalogue brighter than F435W$\sim 19$.

\begin{figure}
	\includegraphics[width=240pt]{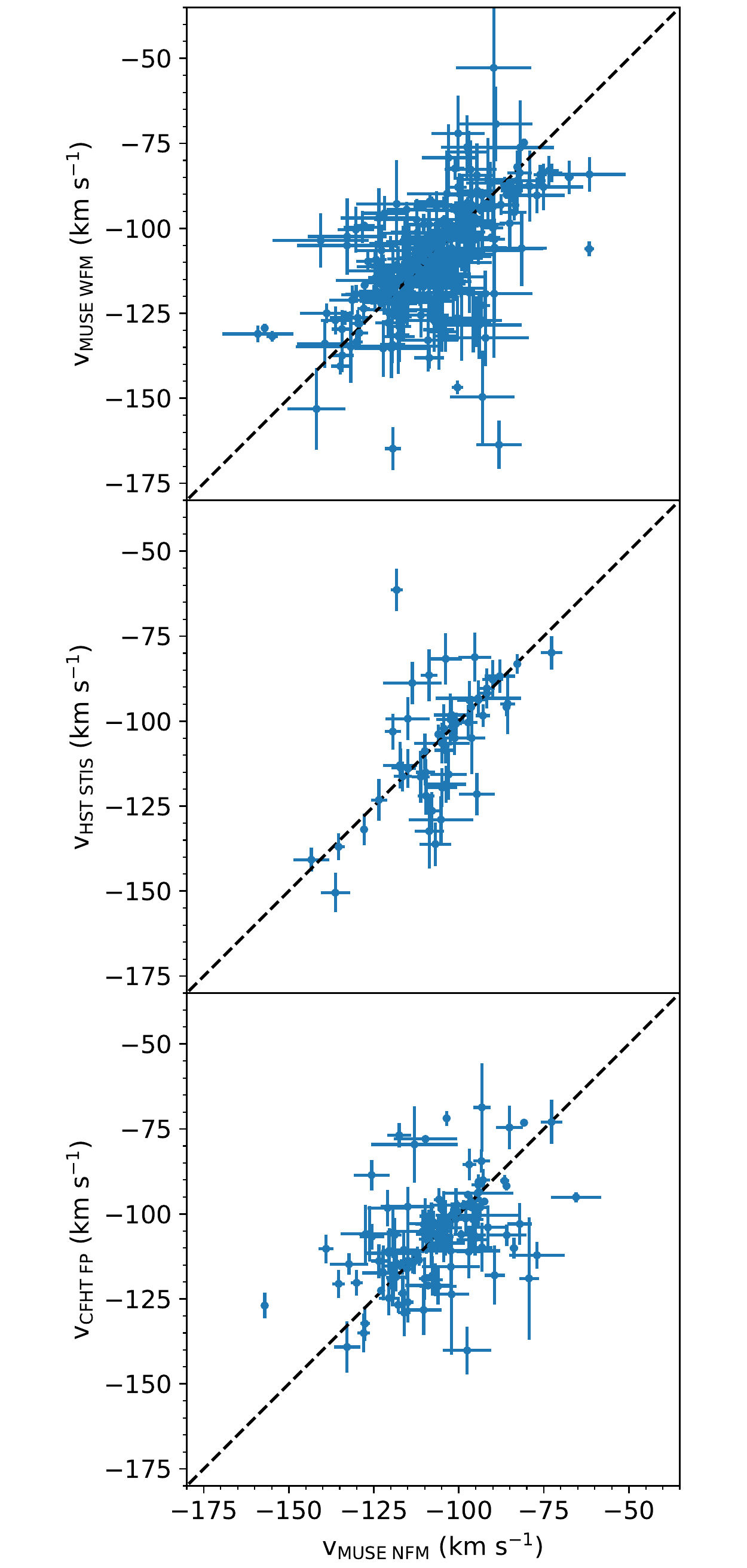}
	\caption{Comparison of the MUSE NFM velocities with velocities from other sources.
		We show the differences between the MUSE NFM velocities and the MUSE WFM velocities (top), the HST STIS velocities (middle) and the CFHT FP velocities (bottom) as a function of the MUSE NFM velocities for the stars in common between the studies.
		The scatter between studies is larger than the uncertainties.}
	\label{fig:velocity_diffs}
\end{figure}

In Figure \ref{fig:velocity_diffs} we show a comparison of the MUSE NFM velocities and the other velocities.
When combining the velocity catalogues we applied shifts to each of the other catalogues such that the median velocities of stars in common were the same as the median for the same stars in the MUSE NFM sample. 
For the 312 stars in common between the MUSE WFM and MUSE NFM datasets, 68 \% of the velocity differences lie between
$-11.5$ and $8.9$ km s$^{-1}$, for the 54 stars in common the HST STIS and MUSE NFM, 68 \% of the velocity differences lie between
$-12.6$ and $3.8$ km s$^{-1}$, and for the 126 stars in common the CFHT and MUSE NFM, 68 \% of the velocity differences lie between
$-11.0$ and $9.8$ km s$^{-1}$.
The widths of the distributions of the velocity differences are roughly two times larger than would be expected from the uncertainties.
The median uncertainties on the MUSE NFM velocities are similar to those of the MUSE WFM velocities (3.5 versus 3.2 km s$^{-1}$) but smaller than the HST STIS (2.8 versus 5.5 km s$^{-1}$) or the CFHT FP (2.0 versus 3.8 km s$^{-1}$) velocity uncertainties for stars in common between the studies.

To check whether the uncertainties are systematically larger or smaller for one sample or another, we used the same Markov chain Monte Carlo kinematics analysis presented in Section \ref{sec:kinematics} to fit the median velocity and the velocity dispersion for the stars in common between the other velocity source and the MUSE NFM using both the NFM velocities and the other sources for each of the MUSE WFM, HST STIS and CFHT FP datasets.
In all cases the fitted kinematics agree within the uncertainties suggesting that the uncertainties of none of the samples are systematically different.

We expect that stars in short period binaries would have radial velocities that would vary between the different studies, inflating the differences in velocities beyond what is expected from their uncertainties.
The binary fraction is also typically higher in the centres of GCs  than in their outskirts \citep[e.g.][]
{2012A&A...540A..16M, 2015ApJ...807...32J,2019A&A...632A...3G}.

\section{Inner kinematics of M15}
\label{sec:kinematics}

\begin{figure}
	\includegraphics[width=240pt]{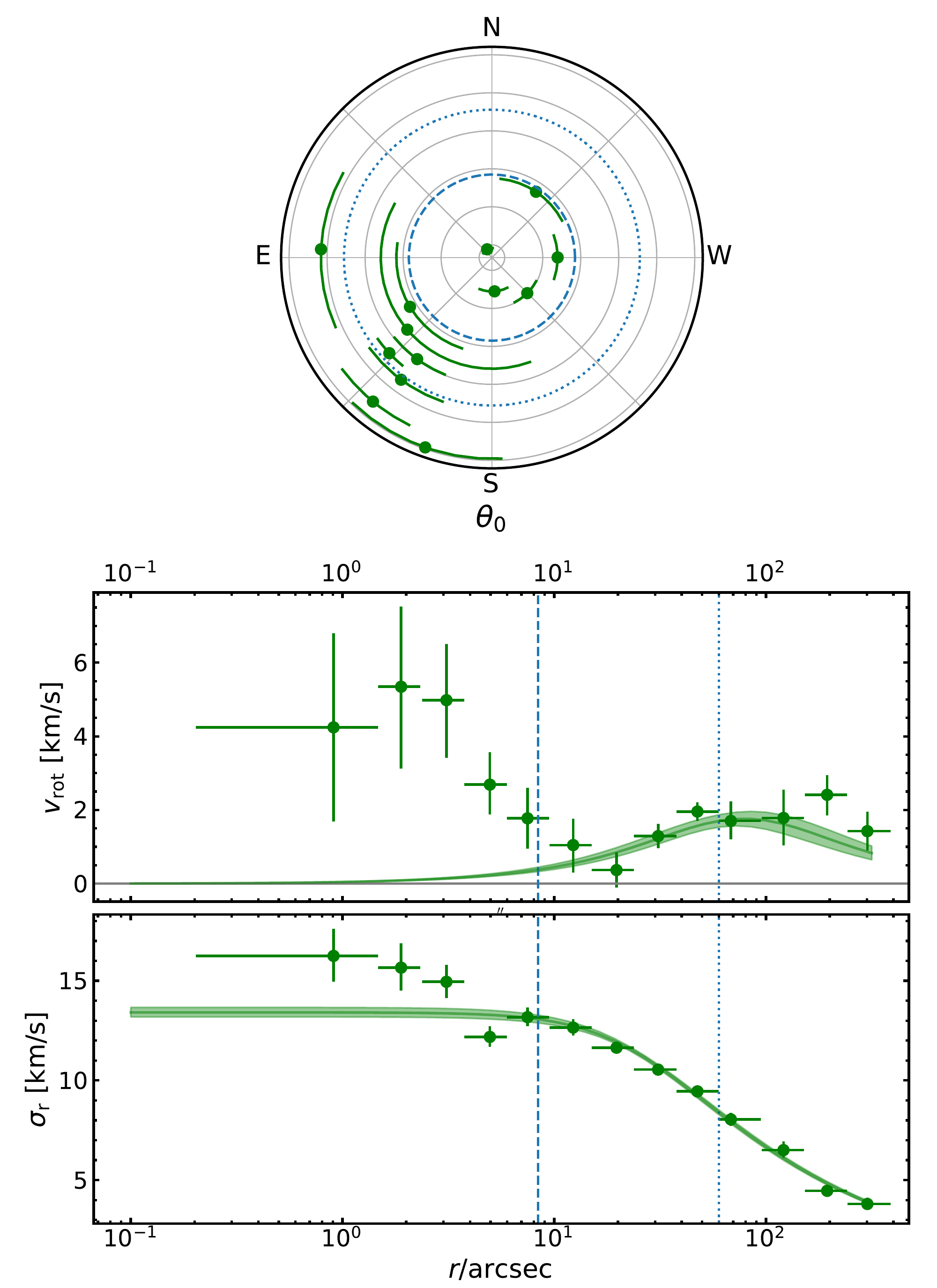}
	\caption{Stellar kinematics of M15.
	The panels show the position angle (top), rotation velocity (middle) and velocity dispersion (bottom), all as a function of clustercentric radius.
	The solid lines show the results of fitting our unbinned data with simple analytical models, while the green circles show the results from radial binning of our dataset.
	The blue dashed and dotted lines show the core and half light radii respectively from the 2010 edition of the \citet{1996AJ....112.1487H} catalogue.
	The MUSE NFM data allows the inner kinematics to be studied in great detail, revealing the strong twist in rotation in the centre of M15 and a double peaked rotation curve that rises towards the centre.}
\label{fig:m15_kinematics_profiles}

\end{figure}

We analysed the spatial resolved kinematics in a similar manner as in \citet{2018MNRAS.473.5591K,2020MNRAS.492..966K}, by deriving 1D radial profiles and 2D Voronoi maps of the first and second moments of the line-of-sight velocity distribution. The radial profiles are depicted in Fig.~\ref{fig:m15_kinematics_profiles}. The continuous curves were obtained by describing the unbinned data using analytical models. More specific, we adopted the rotation profile expected for systems that underwent violent relaxation \citep[][eq.~1 in \citealt{2020MNRAS.492..966K}]{1967MNRAS.136..101L,1973ApJ...186..481G}, while a \citet{1911MNRAS..71..460P} profile \citep[eq.~2 in][] {2020MNRAS.492..966K} was adopted for the dispersion. The model parameters were obtained using a maximum-likelihood approach, where we used the Markov chain Monte Carlo code \textsc{emcee} \citep{2013PASP..125..306F} to sample their posterior distributions. Our model also included a background component, characterised by a Gaussian velocity distribution with mean velocity $v_{\rm back}$ and dispersion $\sigma_{\rm back}$ and a surface stellar density $f_{\rm back}$. Inclusion of the background component allowed us to determine a posteriori cluster membership probabilities for all stars in our sample, via

\begin{equation*}
p_{i}=\left[1+\frac{f_{back}}{1-f_{back}}\sqrt{\frac{\sigma_{back}^{2}+\epsilon_{i}^{2}}{\sigma^{2}+\epsilon_{i}^{2}}} \exp{\left(\frac{(v_{i}-v_{sys})^{2}}{2(\sigma^{2}+\epsilon_{i}^{2})}-\frac{(v_{i}-v_{back})^{2}}{2(\sigma_{back}^{2}+\epsilon_{i}^{2})}\right)}\right]^{-1}\,.
\end{equation*}
where $v_{i}$ and $\epsilon_{i}$ are the radial velocity of the $i$-th star and its associated uncertainty and $\sigma$ is the velocity dispersion of the bin containing it. We do not try to identify or remove the contribution of binary stars.

We also analysed the kinematics in radial bins that contained at least 100 stars and covered at least a radial range of 0.2 dex in log radius. From this, we excluded all stars with membership probabilities $p<0.5$. Further, we fixed the mean velocity in each bin to the value of $v_{\rm sys}=-106.5\pm0.1\,{\rm km\,s^{-1}}$ that was determined in the previous step. Hence, in each bin, we optimized three parameters: the rotation amplitude, the position angle of the rotation axis, and the velocity dispersion. To do so, we used the same maximum-likelihood approach introduced above. The values of the three parameters in all radial bins are included in Fig.~\ref{fig:m15_kinematics_profiles}.

\begin{figure*}
    \centering
    \includegraphics[width=504pt]{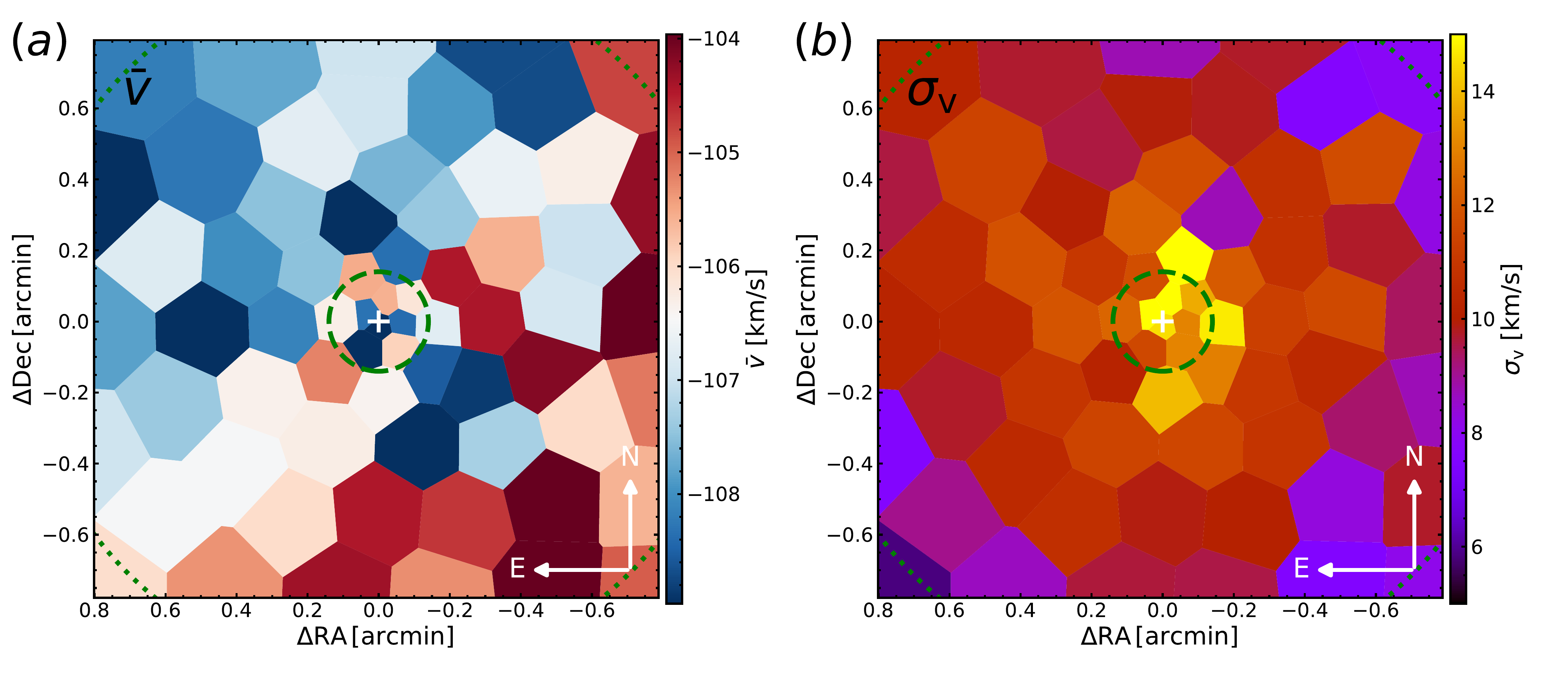}
    \caption{Voronoi maps of the line-of-sight heliocentric velocity (left) and velocity dispersion (right) for the centre of M15.
    Each bin contains $\sim 100$ stars.
    The colour scale of the left panel is centred on the cluster systemic velocity of $-106.5~{\rm km\,s^{-1}}$.
    The solid green circle shows the core radius while the dashed green line shows the half light radius.
    A clear 90 degree twist is seen in the rotation field between the core radius and the half light radius.
    However the velocity field is more complicated than a twisted disc with the negative velocities below and to the left of the centre not having corresponding features above and to the right of the centre that would be expected if the field is symmetric.
    The velocity dispersion map shows that the peak is extended in a north-south direction.}
    \label{fig:m15_voronoi_map}
\end{figure*}

Second, we used the \citet{2003MNRAS.342..345C} code to create Voronoi maps of the kinematics of M15 with an average number of 100 stars per bin. In each bin, we used a simple Nelder-Mead optimization in order to obtain the mean velocity and the velocity dispersion. Stars with membership probabilities $p<0.5$ were again excluded from the calculation. The resulting Voronoi maps of the mean velocity and the velocity dispersion are shown in Fig.~\ref{fig:m15_voronoi_map}.

We see clear evidence for a kinematically distinct core in M15 in both the radial profiles and the Voronoi bins.
While at large radii, the measured rotation profile is largely consistent with the analytic profile (cf. Fig.~\ref{fig:m15_kinematics_profiles}, upper left panel), this is not true for radii $r\lesssim5\arcsec$, where the rotation measurements obtained via radially binning the data diverge from the fitted profile.
Rather than showing a single peak, the rotation amplitude shows a minimum at $\sim 20$ arcsec with two peaks at $\sim 2$ and $\sim 100$ arcsec.
Since the analytic profile is fit to the individual stars and not the binned data, the profile is a better fit to the bins at large radii due to the larger number of stars in this range.
Furthermore, the axis of rotation clearly twists by $\sim 90$ deg at roughly 10 arcsec of the centre (cf. Fig.~\ref{fig:m15_kinematics_profiles}, right panel).
This twist has been previously identified by \citet{2006ApJ...641..852V}, \citet{2013ApJ...772...67B} and \citet{2018MNRAS.473.5591K}.
Another twist may be present even closer to the centre with the inner-most bin showing a rotation axis $\sim 180$ deg different from the next further out bin.

The Voronoi map shows a velocity field more complex and asymmetric than expected from a simple twisted disc model.
For example, a region with negative velocity is seen south-west of the centre that does not have a corresponding region with positive velocity to the north-east.
The velocity dispersion map shows an elongation of the central velocity peak in the north-south direction such that the velocity dispersion north and south of the centre is higher than east or west of the centre at the same distance.

\section{Discussion}
\label{sec:discussion}

Kinematically distinct cores (KDCs) are found in the centres of some massive early-type galaxies (e.g. 19 out 260 galaxies in the ATLAS$^{\text{3D}}$ sample, \citealt{2011MNRAS.414.2923K}).
In galaxies, KDCs are usually thought to form due to mergers \citep[e.g.][]{1984ApJ...287..577K, 1988ApJ...327L..55F, 1990ApJ...361..381B, 1991Natur.354..210H, 2010ApJ...723..818H, 2011MNRAS.416.1654B}.
However, mergers of GC are thought to be relatively rare \citep{2018A&A...620A.154K}. 
We note that M15 does not show any spread in Fe \citep{2009A&A...508..695C} which might be expected if M15 is a merger remnant.

Alternatively, a KDC may be due to the projection effects of a triaxial system with a different distribution of orbits in the centre than in the outskirts \citep{1991AJ....102..882S, 2008MNRAS.385..647V}.
Core collapse could conceivably cause the orbits in the centre of the cluster to be different than in the outskirts.

\citet{2018MNRAS.473.5591K} found a correlation between rotation and relaxation times at the half light radii, suggesting that GCs loose angular momentum via two-body interactions as they dynamically evolve \citep[see][for a theoretical modeling of this process]{2017MNRAS.469..683T}.
Due to their significantly larger sizes, in galaxies two-body relaxation timescales are longer than the age of the Universe so KDCs are expected to persist in them till the present day.
Due to the short two-body relaxation time scale of the core of M15 ($< 100$ Myr, \citealt{1996AJ....112.1487H}), any ordered rotation should be quickly erased.
Thus it is unlikely that the complex kinematics of centre M15 are a relic of its formation.

\citet{2019ApJ...876...87B} used the same HST ACS/HRC imaging used in this work to study the blue straggler population of M15.
They found evidence of multiple generations of blue stragglers formed by stellar collisions.
Beccari et al. suggest that these generations of blue stragglers formed during brief periods of increased stellar density caused by core collapse and by the oscillations that occur after core collapse \citep[e.g.][]{1984MNRAS.208..493B, 1987ApJ...313..576G, 1989ApJ...342..814C, 1991PASJ...43..589T, 2012MNRAS.425.2493B}.
The complicated velocity field of the centre M15 may also be due to these post core collapse oscillations.
The post core collapse model of \citet{1992ApJ...392...86G} reproduces the observed large core radius, cuspy light profile and high central velocity peak of M15.
\citet{1992ApJ...392...86G} also find that the post core collapse oscillations only produce significant temporal variations in surface brightness and in kinematics within $\sim 1$ arcsec.
This would neatly explain how the central kinematics of M15 are disconnected from the outer kinematics.
Detailed dynamical modelling will be required to test whether the observed kinematics can be explained by post core collapse oscillations.

\citet{2005MNRAS.364.1315M} propose that a central binary with massive components (i.e. two massive stellar-mass BHs $\gtrsim20\,\msun$ or two IMBHs) could transfer angular momentum and energy to nearby stars, causing them to co-rotate with the binary and have high velocity dispersion. These predictions are very similar to what we see in the inner few arc seconds and would also naturally explain the misalignment between the inner and outer rotation, but the collapsed core argues strongly against the presence of BH binaries of any mass \citep{2007MNRAS.379...93H, 2007PASJ...59L..11H,2013MNRAS.432.2779B}. It could be that M15 only recently ejected its last BH binary, and still shows features of the co-rotation imprinted by this binary. For this to work, angular momentum needs to be retained during core collapse and some time thereafter. Alternatively, the mechanism proposed by \citet{2005MNRAS.364.1315M} also works with lower mass binaries (e.g. a recently formed binary with non-BH components, such as neutrons stars or white dwarfs).  Dynamical modelling is required to test these ideas and asses whether they can explain the observations.

We note that \citet{2017ApJ...841..114B} discussed the case of another cluster, M53, which also exhibits strong internal rotation in its central regions with a complex radial profile and a radial variation of the position angle of the rotation axis. Their data however do not probe the innermost few arcseconds that we are able to study in M15 with MUSE, and M53 is not a core collapse cluster.
\citet{2018MNRAS.473.5591K} also found evidence for enhanced rotation in the centre and twists in the position angle with radius in five of their 25 GCs (NGC 362, M5, M10, M15 and M62).
All five of these GCs are relatively compact (King profile concentrations $c > 1.7$) but only three (NGC 362, M15 and M62) are listed as possible core collapse GCs in the 2010 edition of the \citet{1996AJ....112.1487H} catalogue.

\section{Conclusion}

Using the new narrow field mode of MUSE we have studied the central kinematics of the core collapse GC M15 in detail.
The combination of AO delivering HST-like spatial resolution with MUSE's large field-of-view and analysis techniques to extract spectra from crowded fields (i.e. PampelMuse) allows for a significantly larger sample of stars (864) with radial velocities compared to earlier studies with AO \citep[104 stars]{2000AJ....119.1268G} and HST \citep[64 stars]{2002AJ....124.3270G}.
The improvements in instrumentation and analysis techniques over the last two decades are highlighted by the fact our significantly larger sample required a significantly smaller amount of telescope time (1.3 hours of exposure time) compared to previous studies (6.5 hours for \citealt{2000AJ....119.1268G}; 25 HST orbits or 16.2 hours of exposure time for \citealt{2002AJ....124.3270G}).

We see clear evidence for complex central kinematics that are distinct from those of the outer parts of the cluster. 
Due to the short two-body relaxation time scale, this KDC is unlikely to have survived from the formation of M15.
Nor is it likely that this KDC shares a similar merger origin to the much larger KDCs observed in some massive early-type galaxies.
In a future work we will use dynamical models to fit this kinematic data set. This will allow us to explore possible hypotheses for the peculiar kinematics of this cluster, including the transfer of angular momentum from black hole binaries and post core-collapse oscillations. The dynamical models will further allow us to place strong constraints on the presence of any compact system of stellar remnants \citep[e.g.][]{2018MNRAS.473.4832G, 2019MNRAS.483.1400H, 2020MNRAS.491..113H}.

\section*{Acknowledgements}
We thank the anonymous referee for their helpful suggestions which improved the manuscript.
CU acknowledges the support of the Swedish Research Council, Vetenskapsr{\aa}det.
SK gratefully acknowledges funding from UKRI in the form of a Future Leaders Fellowship (grant no. MR/T022868/1) and financial support from the European Research Council (ERC-CoG-646928, Multi-Pop). MG acknowledges support from the Ministry of Science and Innovation through a Europa Excelencia grant (EUR2020-112157). VHB acknowledges the support of the Natural Sciences and Engineering Research Council of Canada (NSERC) through grant RGPIN-2020-05990. EB acknowledges financial support from a Vici grant by the Netherlands Organisation for Scientific Research (NWO).

Based on observations collected at the European Organisation for Astronomical Research in the Southern Hemisphere under ESO programme 60.A-9492(A).

This work made use of \textsc{numpy} \citep{numpy}, \textsc{scipy} \citep{scipy}, \textsc{matplotlib} \citep{matplotlib} and \textsc{corner} \citep{corner} as well as \textsc{astropy}, a community-developed core Python package for astronomy \citep{2013A&A...558A..33A}.

\section*{Data availability}
The raw MUSE data are available from the ESO archive (\url{https://archive.eso.org/eso/eso_archive_main.html}).
The reduced data underlying this work is available from the authors upon reasonable request.

\bibliographystyle{mnras}
\bibliography{bib}{}

\begin{thebibliography}{}
\makeatletter
\relax
\def\mn@urlcharsother{\let\do\@makeother \do\$\do\&\do\#\do\^\do\_\do\%\do\~}
\def\mn@doi{\begingroup\mn@urlcharsother \@ifnextchar [ {\mn@doi@}
  {\mn@doi@[]}}
\def\mn@doi@[#1]#2{\def\@tempa{#1}\ifx\@tempa\@empty \href
  {http://dx.doi.org/#2} {doi:#2}\else \href {http://dx.doi.org/#2} {#1}\fi
  \endgroup}
\def\mn@eprint#1#2{\mn@eprint@#1:#2::\@nil}
\def\mn@eprint@arXiv#1{\href {http://arxiv.org/abs/#1} {{\tt arXiv:#1}}}
\def\mn@eprint@dblp#1{\href {http://dblp.uni-trier.de/rec/bibtex/#1.xml}
  {dblp:#1}}
\def\mn@eprint@#1:#2:#3:#4\@nil{\def\@tempa {#1}\def\@tempb {#2}\def\@tempc
  {#3}\ifx \@tempc \@empty \let \@tempc \@tempb \let \@tempb \@tempa \fi \ifx
  \@tempb \@empty \def\@tempb {arXiv}\fi \@ifundefined
  {mn@eprint@\@tempb}{\@tempb:\@tempc}{\expandafter \expandafter \csname
  mn@eprint@\@tempb\endcsname \expandafter{\@tempc}}}

\bibitem[\protect\citeauthoryear{{Abolfathi} et~al.,}{{Abolfathi}
  et~al.}{2018}]{2018ApJS..235...42A}
{Abolfathi} B.,  et~al., 2018, \mn@doi [\apjs] {10.3847/1538-4365/aa9e8a},
  \href {https://ui.adsabs.harvard.edu/abs/2018ApJS..235...42A} {235, 42}

\bibitem[\protect\citeauthoryear{{Arsenault} et~al.,}{{Arsenault}
  et~al.}{2012}]{2012SPIE.8447E..0JA}
{Arsenault} R.,  et~al., 2012, {ESO adaptive optics facility progress report}.
p. 84470J, \mn@doi{10.1117/12.926074}

\bibitem[\protect\citeauthoryear{{Astropy Collaboration} et~al.,}{{Astropy
  Collaboration} et~al.}{2013}]{2013A&A...558A..33A}
{Astropy Collaboration} et~al., 2013, \mn@doi [\aap]
  {10.1051/0004-6361/201322068}, \href
  {http://adsabs.harvard.edu/abs/2013A%26A...558A..33A} {558, A33}

\bibitem[\protect\citeauthoryear{{Bacon} et~al.,}{{Bacon}
  et~al.}{2010}]{2010SPIE.7735E..08B}
{Bacon} R.,  et~al., 2010, {The MUSE second-generation VLT instrument}.
p. 773508, \mn@doi{10.1117/12.856027}

\bibitem[\protect\citeauthoryear{{Balcells} \& {Quinn}}{{Balcells} \&
  {Quinn}}{1990}]{1990ApJ...361..381B}
{Balcells} M.,  {Quinn} P.~J.,  1990, \mn@doi [\apj] {10.1086/169204}, \href
  {https://ui.adsabs.harvard.edu/abs/1990ApJ...361..381B} {361, 381}

\bibitem[\protect\citeauthoryear{{Baumgardt} \& {Hilker}}{{Baumgardt} \&
  {Hilker}}{2018}]{2018MNRAS.478.1520B}
{Baumgardt} H.,  {Hilker} M.,  2018, \mn@doi [\mnras] {10.1093/mnras/sty1057},
  \href {https://ui.adsabs.harvard.edu/#abs/2018MNRAS.478.1520B} {478, 1520}

\bibitem[\protect\citeauthoryear{{Baumgardt}, {Hut}, {Makino}, {McMillan}  \&
  {Portegies Zwart}}{{Baumgardt} et~al.}{2003}]{2003ApJ...582L..21B}
{Baumgardt} H.,  {Hut} P.,  {Makino} J.,  {McMillan} S.,   {Portegies Zwart}
  S.,  2003, \mn@doi [\apjl] {10.1086/367537}, \href
  {https://ui.adsabs.harvard.edu/abs/2003ApJ...582L..21B} {582, L21}

\bibitem[\protect\citeauthoryear{{Beccari} et~al.,}{{Beccari}
  et~al.}{2019}]{2019ApJ...876...87B}
{Beccari} G.,  et~al., 2019, \mn@doi [\apj] {10.3847/1538-4357/ab13a4}, \href
  {https://ui.adsabs.harvard.edu/abs/2019ApJ...876...87B} {876, 87}

\bibitem[\protect\citeauthoryear{{Bellini} et~al.,}{{Bellini}
  et~al.}{2014}]{2014ApJ...797..115B}
{Bellini} A.,  et~al., 2014, \mn@doi [\apj] {10.1088/0004-637X/797/2/115},
  \href {https://ui.adsabs.harvard.edu/abs/2014ApJ...797..115B} {797, 115}

\bibitem[\protect\citeauthoryear{{Bettwieser} \& {Sugimoto}}{{Bettwieser} \&
  {Sugimoto}}{1984}]{1984MNRAS.208..493B}
{Bettwieser} E.,  {Sugimoto} D.,  1984, \mn@doi [\mnras]
  {10.1093/mnras/208.3.493}, \href
  {https://ui.adsabs.harvard.edu/abs/1984MNRAS.208..493B} {208, 493}

\bibitem[\protect\citeauthoryear{{Bianchini}, {Varri}, {Bertin}  \&
  {Zocchi}}{{Bianchini} et~al.}{2013}]{2013ApJ...772...67B}
{Bianchini} P.,  {Varri} A.~L.,  {Bertin} G.,   {Zocchi} A.,  2013, \mn@doi
  [\apj] {10.1088/0004-637X/772/1/67}, \href
  {https://ui.adsabs.harvard.edu/abs/2013ApJ...772...67B} {772, 67}

\bibitem[\protect\citeauthoryear{{Bianchini}, {van der Marel}, {del Pino},
  {Watkins}, {Bellini}, {Fardal}, {Libralato}  \& {Sills}}{{Bianchini}
  et~al.}{2018}]{2018MNRAS.481.2125B}
{Bianchini} P.,  {van der Marel} R.~P.,  {del Pino} A.,  {Watkins} L.~L.,
  {Bellini} A.,  {Fardal} M.~A.,  {Libralato} M.,   {Sills} A.,  2018, \mn@doi
  [\mnras] {10.1093/mnras/sty2365}, \href
  {https://ui.adsabs.harvard.edu/abs/2018MNRAS.481.2125B} {481, 2125}

\bibitem[\protect\citeauthoryear{{Boberg}, {Vesperini}, {Friel}, {Tiongco}  \&
  {Varri}}{{Boberg} et~al.}{2017}]{2017ApJ...841..114B}
{Boberg} O.~M.,  {Vesperini} E.,  {Friel} E.~D.,  {Tiongco} M.~A.,   {Varri}
  A.~L.,  2017, \mn@doi [\apj] {10.3847/1538-4357/aa7070}, \href
  {https://ui.adsabs.harvard.edu/abs/2017ApJ...841..114B} {841, 114}

\bibitem[\protect\citeauthoryear{{Bois} et~al.,}{{Bois}
  et~al.}{2011}]{2011MNRAS.416.1654B}
{Bois} M.,  et~al., 2011, \mn@doi [\mnras] {10.1111/j.1365-2966.2011.19113.x},
  \href {https://ui.adsabs.harvard.edu/abs/2011MNRAS.416.1654B} {416, 1654}

\bibitem[\protect\citeauthoryear{{Breen} \& {Heggie}}{{Breen} \&
  {Heggie}}{2012}]{2012MNRAS.425.2493B}
{Breen} P.~G.,  {Heggie} D.~C.,  2012, \mn@doi [\mnras]
  {10.1111/j.1365-2966.2012.21688.x}, \href
  {https://ui.adsabs.harvard.edu/abs/2012MNRAS.425.2493B} {425, 2493}

\bibitem[\protect\citeauthoryear{{Breen} \& {Heggie}}{{Breen} \&
  {Heggie}}{2013}]{2013MNRAS.432.2779B}
{Breen} P.~G.,  {Heggie} D.~C.,  2013, \mn@doi [\mnras] {10.1093/mnras/stt628},
  \href {http://adsabs.harvard.edu/abs/2013MNRAS.432.2779B} {432, 2779}

\bibitem[\protect\citeauthoryear{{Cappellari} \& {Copin}}{{Cappellari} \&
  {Copin}}{2003}]{2003MNRAS.342..345C}
{Cappellari} M.,  {Copin} Y.,  2003, \mn@doi [\mnras]
  {10.1046/j.1365-8711.2003.06541.x}, \href
  {https://ui.adsabs.harvard.edu/abs/2003MNRAS.342..345C} {342, 345}

\bibitem[\protect\citeauthoryear{{Carretta}, {Bragaglia}, {Gratton}, {D'Orazi}
  \& {Lucatello}}{{Carretta} et~al.}{2009}]{2009A&A...508..695C}
{Carretta} E.,  {Bragaglia} A.,  {Gratton} R.,  {D'Orazi} V.,   {Lucatello} S.,
   2009, \mn@doi [\aap] {10.1051/0004-6361/200913003}, \href
  {http://adsabs.harvard.edu/abs/2009A%26A...508..695C} {508, 695}

\bibitem[\protect\citeauthoryear{{Choi}, {Dotter}, {Conroy}, {Cantiello},
  {Paxton}  \& {Johnson}}{{Choi} et~al.}{2016}]{2016ApJ...823..102C}
{Choi} J.,  {Dotter} A.,  {Conroy} C.,  {Cantiello} M.,  {Paxton} B.,
  {Johnson} B.~D.,  2016, \mn@doi [\apj] {10.3847/0004-637X/823/2/102}, \href
  {https://ui.adsabs.harvard.edu/abs/2016ApJ...823..102C} {823, 102}

\bibitem[\protect\citeauthoryear{{Cohn}, {Hut}  \& {Wise}}{{Cohn}
  et~al.}{1989}]{1989ApJ...342..814C}
{Cohn} H.,  {Hut} P.,   {Wise} M.,  1989, \mn@doi [\apj] {10.1086/167638},
  \href {https://ui.adsabs.harvard.edu/abs/1989ApJ...342..814C} {342, 814}

\bibitem[\protect\citeauthoryear{{Dalessandro} et~al.,}{{Dalessandro}
  et~al.}{2016}]{2016ApJ...833..111D}
{Dalessandro} E.,  et~al., 2016, \mn@doi [\apj] {10.3847/1538-4357/833/1/111},
  \href {https://ui.adsabs.harvard.edu/abs/2016ApJ...833..111D} {833, 111}

\bibitem[\protect\citeauthoryear{{Dalgleish} et~al.,}{{Dalgleish}
  et~al.}{2020}]{2020MNRAS.492.3859D}
{Dalgleish} H.,  et~al., 2020, \mn@doi [\mnras] {10.1093/mnras/staa091}, \href
  {https://ui.adsabs.harvard.edu/abs/2020MNRAS.492.3859D} {492, 3859}

\bibitem[\protect\citeauthoryear{{Djorgovski} \& {King}}{{Djorgovski} \&
  {King}}{1986}]{1986ApJ...305L..61D}
{Djorgovski} S.,  {King} I.~R.,  1986, \mn@doi [\apjl] {10.1086/184685}, \href
  {https://ui.adsabs.harvard.edu/abs/1986ApJ...305L..61D} {305, L61}

\bibitem[\protect\citeauthoryear{{Dotter} et~al.,}{{Dotter}
  et~al.}{2010}]{2010ApJ...708..698D}
{Dotter} A.,  et~al., 2010, \mn@doi [\apj] {10.1088/0004-637X/708/1/698}, \href
  {http://adsabs.harvard.edu/abs/2010ApJ...708..698D} {708, 698}

\bibitem[\protect\citeauthoryear{{Ferraro} et~al.,}{{Ferraro}
  et~al.}{2018}]{2018ApJ...860...50F}
{Ferraro} F.~R.,  et~al., 2018, \mn@doi [\apj] {10.3847/1538-4357/aabe2f},
  \href {https://ui.adsabs.harvard.edu/abs/2018ApJ...860...50F} {860, 50}

\bibitem[\protect\citeauthoryear{Foreman-Mackey}{Foreman-Mackey}{2016}]{corner}
Foreman-Mackey D.,  2016, \mn@doi [The Journal of Open Source Software]
  {10.21105/joss.00024}, 24

\bibitem[\protect\citeauthoryear{{Foreman-Mackey}, {Hogg}, {Lang}  \&
  {Goodman}}{{Foreman-Mackey} et~al.}{2013}]{2013PASP..125..306F}
{Foreman-Mackey} D.,  {Hogg} D.~W.,  {Lang} D.,   {Goodman} J.,  2013, \mn@doi
  [\pasp] {10.1086/670067}, \href
  {https://ui.adsabs.harvard.edu/abs/2013PASP..125..306F} {125, 306}

\bibitem[\protect\citeauthoryear{{Franx} \& {Illingworth}}{{Franx} \&
  {Illingworth}}{1988}]{1988ApJ...327L..55F}
{Franx} M.,  {Illingworth} G.~D.,  1988, \mn@doi [\apjl] {10.1086/185139},
  \href {https://ui.adsabs.harvard.edu/abs/1988ApJ...327L..55F} {327, L55}

\bibitem[\protect\citeauthoryear{{Gaia Collaboration} et~al.,}{{Gaia
  Collaboration} et~al.}{2018}]{2018A&A...616A..12G}
{Gaia Collaboration} et~al., 2018, \mn@doi [\aap]
  {10.1051/0004-6361/201832698}, \href
  {https://ui.adsabs.harvard.edu/abs/2018A&A...616A..12G} {616, A12}

\bibitem[\protect\citeauthoryear{{Gebhardt}, {Pryor}, {Williams}  \&
  {Hesser}}{{Gebhardt} et~al.}{1994}]{1994AJ....107.2067G}
{Gebhardt} K.,  {Pryor} C.,  {Williams} T.~B.,   {Hesser} J.~E.,  1994, \mn@doi
  [\aj] {10.1086/117017}, \href
  {https://ui.adsabs.harvard.edu/abs/1994AJ....107.2067G} {107, 2067}

\bibitem[\protect\citeauthoryear{{Gebhardt}, {Pryor}, {Williams}, {Hesser}  \&
  {Stetson}}{{Gebhardt} et~al.}{1997}]{1997AJ....113.1026G}
{Gebhardt} K.,  {Pryor} C.,  {Williams} T.~B.,  {Hesser} J.~E.,   {Stetson}
  P.~B.,  1997, \mn@doi [\aj] {10.1086/118319}, \href
  {https://ui.adsabs.harvard.edu/abs/1997AJ....113.1026G} {113, 1026}

\bibitem[\protect\citeauthoryear{{Gebhardt}, {Pryor}, {O'Connell}, {Williams}
  \& {Hesser}}{{Gebhardt} et~al.}{2000}]{2000AJ....119.1268G}
{Gebhardt} K.,  {Pryor} C.,  {O'Connell} R.~D.,  {Williams} T.~B.,   {Hesser}
  J.~E.,  2000, \mn@doi [\aj] {10.1086/301275}, \href
  {https://ui.adsabs.harvard.edu/abs/2000AJ....119.1268G} {119, 1268}

\bibitem[\protect\citeauthoryear{{Gerssen}, {van der Marel}, {Gebhardt},
  {Guhathakurta}, {Peterson}  \& {Pryor}}{{Gerssen}
  et~al.}{2002}]{2002AJ....124.3270G}
{Gerssen} J.,  {van der Marel} R.~P.,  {Gebhardt} K.,  {Guhathakurta} P.,
  {Peterson} R.~C.,   {Pryor} C.,  2002, \mn@doi [\aj] {10.1086/344584}, \href
  {https://ui.adsabs.harvard.edu/abs/2002AJ....124.3270G} {124, 3270}

\bibitem[\protect\citeauthoryear{{Gieles}, {Balbinot}, {Yaaqib},
  {H{\'e}nault-Brunet}, {Zocchi}, {Peuten}  \& {Jonker}}{{Gieles}
  et~al.}{2018}]{2018MNRAS.473.4832G}
{Gieles} M.,  {Balbinot} E.,  {Yaaqib} R. I.~S.~M.,  {H{\'e}nault-Brunet} V.,
  {Zocchi} A.,  {Peuten} M.,   {Jonker} P.~G.,  2018, \mn@doi [\mnras]
  {10.1093/mnras/stx2694}, \href
  {https://ui.adsabs.harvard.edu/abs/2018MNRAS.473.4832G} {473, 4832}

\bibitem[\protect\citeauthoryear{{Giesers} et~al.,}{{Giesers}
  et~al.}{2019}]{2019A&A...632A...3G}
{Giesers} B.,  et~al., 2019, \mn@doi [\aap] {10.1051/0004-6361/201936203},
  \href {https://ui.adsabs.harvard.edu/abs/2019A&A...632A...3G} {632, A3}

\bibitem[\protect\citeauthoryear{{Goodman}}{{Goodman}}{1987}]{1987ApJ...313..576G}
{Goodman} J.,  1987, \mn@doi [\apj] {10.1086/164998}, \href
  {https://ui.adsabs.harvard.edu/abs/1987ApJ...313..576G} {313, 576}

\bibitem[\protect\citeauthoryear{{Gott}}{{Gott}}{1973}]{1973ApJ...186..481G}
{Gott} III R.~J.,  1973, \mn@doi [\apj] {10.1086/152514}, \href
  {http://adsabs.harvard.edu/abs/1973ApJ...186..481G} {186, 481}

\bibitem[\protect\citeauthoryear{{Grabhorn}, {Cohn}, {Lugger}  \&
  {Murphy}}{{Grabhorn} et~al.}{1992}]{1992ApJ...392...86G}
{Grabhorn} R.~P.,  {Cohn} H.~N.,  {Lugger} P.~M.,   {Murphy} B.~W.,  1992,
  \mn@doi [\apj] {10.1086/171408}, \href
  {https://ui.adsabs.harvard.edu/abs/1992ApJ...392...86G} {392, 86}

\bibitem[\protect\citeauthoryear{{Harris}}{{Harris}}{1996}]{1996AJ....112.1487H}
{Harris} W.~E.,  1996, \mn@doi [\aj] {10.1086/118116}, \href
  {http://adsabs.harvard.edu/abs/1996AJ....112.1487H} {112, 1487}

\bibitem[\protect\citeauthoryear{{Heggie}, {Hut}, {Mineshige}, {Makino}  \&
  {Baumgardt}}{{Heggie} et~al.}{2007}]{2007PASJ...59L..11H}
{Heggie} D.~C.,  {Hut} P.,  {Mineshige} S.,  {Makino} J.,   {Baumgardt} H.,
  2007, \mn@doi [\pasj] {10.1093/pasj/59.3.L11}, \href
  {http://adsabs.harvard.edu/abs/2007PASJ...59L..11H} {59, L11}

\bibitem[\protect\citeauthoryear{{H{\'e}nault-Brunet}, {Gieles}, {Sollima},
  {Watkins}, {Zocchi}, {Claydon}, {Pancino}  \&
  {Baumgardt}}{{H{\'e}nault-Brunet} et~al.}{2019}]{2019MNRAS.483.1400H}
{H{\'e}nault-Brunet} V.,  {Gieles} M.,  {Sollima} A.,  {Watkins} L.~L.,
  {Zocchi} A.,  {Claydon} I.,  {Pancino} E.,   {Baumgardt} H.,  2019, \mn@doi
  [\mnras] {10.1093/mnras/sty3187}, \href
  {https://ui.adsabs.harvard.edu/abs/2019MNRAS.483.1400H} {483, 1400}

\bibitem[\protect\citeauthoryear{{H{\'e}nault-Brunet}, {Gieles}, {Strader},
  {Peuten}, {Balbinot}  \& {Douglas}}{{H{\'e}nault-Brunet}
  et~al.}{2020}]{2020MNRAS.491..113H}
{H{\'e}nault-Brunet} V.,  {Gieles} M.,  {Strader} J.,  {Peuten} M.,  {Balbinot}
  E.,   {Douglas} K.~E.~K.,  2020, \mn@doi [\mnras] {10.1093/mnras/stz2995},
  \href {https://ui.adsabs.harvard.edu/abs/2020MNRAS.491..113H} {491, 113}

\bibitem[\protect\citeauthoryear{{Hernquist} \& {Barnes}}{{Hernquist} \&
  {Barnes}}{1991}]{1991Natur.354..210H}
{Hernquist} L.,  {Barnes} J.~E.,  1991, \mn@doi [\nat] {10.1038/354210a0},
  \href {https://ui.adsabs.harvard.edu/abs/1991Natur.354..210H} {354, 210}

\bibitem[\protect\citeauthoryear{{Hoffman}, {Cox}, {Dutta}  \&
  {Hernquist}}{{Hoffman} et~al.}{2010}]{2010ApJ...723..818H}
{Hoffman} L.,  {Cox} T.~J.,  {Dutta} S.,   {Hernquist} L.,  2010, \mn@doi
  [\apj] {10.1088/0004-637X/723/1/818}, \href
  {https://ui.adsabs.harvard.edu/abs/2010ApJ...723..818H} {723, 818}

\bibitem[\protect\citeauthoryear{Hunter}{Hunter}{2007}]{matplotlib}
Hunter J.~D.,  2007, \mn@doi [Computing In Science \& Engineering]
  {10.1109/MCSE.2007.55}, 9, 90

\bibitem[\protect\citeauthoryear{{Hurley}}{{Hurley}}{2007}]{2007MNRAS.379...93H}
{Hurley} J.~R.,  2007, \mn@doi [\mnras] {10.1111/j.1365-2966.2007.11912.x},
  \href {http://adsabs.harvard.edu/abs/2007MNRAS.379...93H} {379, 93}

\bibitem[\protect\citeauthoryear{{Husser}, {Wende-von Berg}, {Dreizler},
  {Homeier}, {Reiners}, {Barman}  \& {Hauschildt}}{{Husser}
  et~al.}{2013}]{2013A&A...553A...6H}
{Husser} T.~O.,  {Wende-von Berg} S.,  {Dreizler} S.,  {Homeier} D.,  {Reiners}
  A.,  {Barman} T.,   {Hauschildt} P.~H.,  2013, \mn@doi [\aap]
  {10.1051/0004-6361/201219058}, \href
  {https://ui.adsabs.harvard.edu/abs/2013A&A...553A...6H} {553, A6}

\bibitem[\protect\citeauthoryear{{Husser} et~al.,}{{Husser}
  et~al.}{2016}]{2016A&A...588A.148H}
{Husser} T.-O.,  et~al., 2016, \mn@doi [\aap] {10.1051/0004-6361/201526949},
  \href {https://ui.adsabs.harvard.edu/abs/2016A&A...588A.148H} {588, A148}

\bibitem[\protect\citeauthoryear{{Illingworth} \& {King}}{{Illingworth} \&
  {King}}{1977}]{1977ApJ...218L.109I}
{Illingworth} G.,  {King} I.~R.,  1977, \mn@doi [\apjl] {10.1086/182586}, \href
  {https://ui.adsabs.harvard.edu/abs/1977ApJ...218L.109I} {218, L109}

\bibitem[\protect\citeauthoryear{{Ji} \& {Bregman}}{{Ji} \&
  {Bregman}}{2015}]{2015ApJ...807...32J}
{Ji} J.,  {Bregman} J.~N.,  2015, \mn@doi [\apj] {10.1088/0004-637X/807/1/32},
  \href {https://ui.adsabs.harvard.edu/abs/2015ApJ...807...32J} {807, 32}

\bibitem[\protect\citeauthoryear{Jones, Oliphant, Peterson  et~al.}{Jones
  et~al.}{01  }]{scipy}
Jones E.,  Oliphant T.,  Peterson P.,   et~al., 2001--, {SciPy}: Open source
  scientific tools for {Python}, \url {http://www.scipy.org/}

\bibitem[\protect\citeauthoryear{{Kamann}, {Wisotzki}  \& {Roth}}{{Kamann}
  et~al.}{2013}]{2013A&A...549A..71K}
{Kamann} S.,  {Wisotzki} L.,   {Roth} M.~M.,  2013, \mn@doi [\aap]
  {10.1051/0004-6361/201220476}, \href
  {http://adsabs.harvard.edu/abs/2013A%26A...549A..71K} {549, A71}

\bibitem[\protect\citeauthoryear{{Kamann}, {Wisotzki}, {Roth}, {Gerssen},
  {Husser}, {Sandin}  \& {Weilbacher}}{{Kamann}
  et~al.}{2014}]{2014A&A...566A..58K}
{Kamann} S.,  {Wisotzki} L.,  {Roth} M.~M.,  {Gerssen} J.,  {Husser} T.~O.,
  {Sandin} C.,   {Weilbacher} P.,  2014, \mn@doi [\aap]
  {10.1051/0004-6361/201322183}, \href
  {https://ui.adsabs.harvard.edu/abs/2014A&A...566A..58K} {566, A58}

\bibitem[\protect\citeauthoryear{{Kamann} et~al.,}{{Kamann}
  et~al.}{2016}]{2016A&A...588A.149K}
{Kamann} S.,  et~al., 2016, \mn@doi [\aap] {10.1051/0004-6361/201527065}, \href
  {https://ui.adsabs.harvard.edu/abs/2016A&A...588A.149K} {588, A149}

\bibitem[\protect\citeauthoryear{{Kamann} et~al.,}{{Kamann}
  et~al.}{2018a}]{2018MNRAS.473.5591K}
{Kamann} S.,  et~al., 2018a, \mn@doi [\mnras] {10.1093/mnras/stx2719}, \href
  {https://ui.adsabs.harvard.edu/abs/2018MNRAS.473.5591K} {473, 5591}

\bibitem[\protect\citeauthoryear{{Kamann} et~al.,}{{Kamann}
  et~al.}{2018b}]{2018MNRAS.480.1689K}
{Kamann} S.,  et~al., 2018b, \mn@doi [\mnras] {10.1093/mnras/sty1958}, \href
  {https://ui.adsabs.harvard.edu/abs/2018MNRAS.480.1689K} {480, 1689}

\bibitem[\protect\citeauthoryear{{Kamann} et~al.,}{{Kamann}
  et~al.}{2020}]{2020MNRAS.492..966K}
{Kamann} S.,  et~al., 2020, \mn@doi [\mnras] {10.1093/mnras/stz3506}, \href
  {https://ui.adsabs.harvard.edu/abs/2020MNRAS.492..966K} {492, 966}

\bibitem[\protect\citeauthoryear{{Khoperskov}, {Mastrobuono-Battisti}, {Di
  Matteo}  \& {Haywood}}{{Khoperskov} et~al.}{2018}]{2018A&A...620A.154K}
{Khoperskov} S.,  {Mastrobuono-Battisti} A.,  {Di Matteo} P.,   {Haywood} M.,
  2018, \mn@doi [\aap] {10.1051/0004-6361/201833534}, \href
  {https://ui.adsabs.harvard.edu/abs/2018A&A...620A.154K} {620, A154}

\bibitem[\protect\citeauthoryear{{Kimmig}, {Seth}, {Ivans}, {Strader},
  {Caldwell}, {Anderton}  \& {Gregersen}}{{Kimmig}
  et~al.}{2015}]{2015AJ....149...53K}
{Kimmig} B.,  {Seth} A.,  {Ivans} I.~I.,  {Strader} J.,  {Caldwell} N.,
  {Anderton} T.,   {Gregersen} D.,  2015, \mn@doi [\aj]
  {10.1088/0004-6256/149/2/53}, \href
  {https://ui.adsabs.harvard.edu/abs/2015AJ....149...53K} {149, 53}

\bibitem[\protect\citeauthoryear{{Kormendy}}{{Kormendy}}{1984}]{1984ApJ...287..577K}
{Kormendy} J.,  1984, \mn@doi [\apj] {10.1086/162717}, \href
  {https://ui.adsabs.harvard.edu/abs/1984ApJ...287..577K} {287, 577}

\bibitem[\protect\citeauthoryear{{Krajnovi{\'c}} et~al.,}{{Krajnovi{\'c}}
  et~al.}{2011}]{2011MNRAS.414.2923K}
{Krajnovi{\'c}} D.,  et~al., 2011, \mn@doi [\mnras]
  {10.1111/j.1365-2966.2011.18560.x}, \href
  {https://ui.adsabs.harvard.edu/abs/2011MNRAS.414.2923K} {414, 2923}

\bibitem[\protect\citeauthoryear{{Lane}, {Kiss}, {Lewis}, {Ibata}, {Siebert},
  {Bedding}, {Sz{\'e}kely}  \& {Szab{\'o}}}{{Lane}
  et~al.}{2011}]{2011A&A...530A..31L}
{Lane} R.~R.,  {Kiss} L.~L.,  {Lewis} G.~F.,  {Ibata} R.~A.,  {Siebert} A.,
  {Bedding} T.~R.,  {Sz{\'e}kely} P.,   {Szab{\'o}} G.~M.,  2011, \mn@doi
  [\aap] {10.1051/0004-6361/201116660}, \href
  {https://ui.adsabs.harvard.edu/abs/2011A&A...530A..31L} {530, A31}

\bibitem[\protect\citeauthoryear{{Lanzoni} et~al.,}{{Lanzoni}
  et~al.}{2013}]{2013ApJ...769..107L}
{Lanzoni} B.,  et~al., 2013, \mn@doi [\apj] {10.1088/0004-637X/769/2/107},
  \href {https://ui.adsabs.harvard.edu/abs/2013ApJ...769..107L} {769, 107}

\bibitem[\protect\citeauthoryear{{Lynden-Bell}}{{Lynden-Bell}}{1967}]{1967MNRAS.136..101L}
{Lynden-Bell} D.,  1967, \mn@doi [\mnras] {10.1093/mnras/136.1.101}, \href
  {http://adsabs.harvard.edu/abs/1967MNRAS.136..101L} {136, 101}

\bibitem[\protect\citeauthoryear{{Mapelli}, {Colpi}, {Possenti}  \&
  {Sigurdsson}}{{Mapelli} et~al.}{2005}]{2005MNRAS.364.1315M}
{Mapelli} M.,  {Colpi} M.,  {Possenti} A.,   {Sigurdsson} S.,  2005, \mn@doi
  [\mnras] {10.1111/j.1365-2966.2005.09653.x}, \href
  {https://ui.adsabs.harvard.edu/abs/2005MNRAS.364.1315M} {364, 1315}

\bibitem[\protect\citeauthoryear{{Massari} et~al.,}{{Massari}
  et~al.}{2016}]{2016A&A...595L...2M}
{Massari} D.,  et~al., 2016, \mn@doi [\aap] {10.1051/0004-6361/201629336},
  \href {https://ui.adsabs.harvard.edu/abs/2016A&A...595L...2M} {595, L2}

\bibitem[\protect\citeauthoryear{{McNamara}, {Harrison}  \&
  {Anderson}}{{McNamara} et~al.}{2003}]{2003ApJ...595..187M}
{McNamara} B.~J.,  {Harrison} T.~E.,   {Anderson} J.,  2003, \mn@doi [\apj]
  {10.1086/377341}, \href
  {https://ui.adsabs.harvard.edu/abs/2003ApJ...595..187M} {595, 187}

\bibitem[\protect\citeauthoryear{{Milone} et~al.,}{{Milone}
  et~al.}{2012}]{2012A&A...540A..16M}
{Milone} A.~P.,  et~al., 2012, \mn@doi [\aap] {10.1051/0004-6361/201016384},
  \href {https://ui.adsabs.harvard.edu/abs/2012A&A...540A..16M} {540, A16}

\bibitem[\protect\citeauthoryear{{Murphy}, {Cohn}  \& {Lugger}}{{Murphy}
  et~al.}{2011}]{2011ApJ...732...67M}
{Murphy} B.~W.,  {Cohn} H.~N.,   {Lugger} P.~M.,  2011, \mn@doi [\apj]
  {10.1088/0004-637X/732/2/67}, \href
  {https://ui.adsabs.harvard.edu/abs/2011ApJ...732...67M} {732, 67}

\bibitem[\protect\citeauthoryear{{Newell}, {Da Costa}  \& {Norris}}{{Newell}
  et~al.}{1976}]{1976ApJ...208L..55N}
{Newell} B.,  {Da Costa} G.~S.,   {Norris} J.,  1976, \mn@doi [\apjl]
  {10.1086/182232}, \href
  {https://ui.adsabs.harvard.edu/abs/1976ApJ...208L..55N} {208, L55}

\bibitem[\protect\citeauthoryear{{Pasquini} et~al.,}{{Pasquini}
  et~al.}{2002}]{2002Msngr.110....1P}
{Pasquini} L.,  et~al., 2002, The Messenger, \href
  {https://ui.adsabs.harvard.edu/abs/2002Msngr.110....1P} {110, 1}

\bibitem[\protect\citeauthoryear{{Peterson}, {Seitzer}  \&
  {Cudworth}}{{Peterson} et~al.}{1989}]{1989ApJ...347..251P}
{Peterson} R.~C.,  {Seitzer} P.,   {Cudworth} K.~M.,  1989, \mn@doi [\apj]
  {10.1086/168114}, \href
  {https://ui.adsabs.harvard.edu/abs/1989ApJ...347..251P} {347, 251}

\bibitem[\protect\citeauthoryear{{Plummer}}{{Plummer}}{1911}]{1911MNRAS..71..460P}
{Plummer} H.~C.,  1911, \mn@doi [\mnras] {10.1093/mnras/71.5.460}, \href
  {http://adsabs.harvard.edu/abs/1911MNRAS..71..460P} {71, 460}

\bibitem[\protect\citeauthoryear{{Rigaut} et~al.,}{{Rigaut}
  et~al.}{1998}]{1998PASP..110..152R}
{Rigaut} F.,  et~al., 1998, \mn@doi [\pasp] {10.1086/316126}, \href
  {https://ui.adsabs.harvard.edu/abs/1998PASP..110..152R} {110, 152}

\bibitem[\protect\citeauthoryear{{Sarajedini} et~al.,}{{Sarajedini}
  et~al.}{2007}]{2007AJ....133.1658S}
{Sarajedini} A.,  et~al., 2007, \mn@doi [\aj] {10.1086/511979}, \href
  {http://adsabs.harvard.edu/abs/2007AJ....133.1658S} {133, 1658}

\bibitem[\protect\citeauthoryear{{Sollima}, {Baumgardt}  \& {Hilker}}{{Sollima}
  et~al.}{2019}]{2019MNRAS.485.1460S}
{Sollima} A.,  {Baumgardt} H.,   {Hilker} M.,  2019, \mn@doi [\mnras]
  {10.1093/mnras/stz505}, \href
  {https://ui.adsabs.harvard.edu/abs/2019MNRAS.485.1460S} {485, 1460}

\bibitem[\protect\citeauthoryear{{Statler}}{{Statler}}{1991}]{1991AJ....102..882S}
{Statler} T.~S.,  1991, \mn@doi [\aj] {10.1086/115919}, \href
  {https://ui.adsabs.harvard.edu/abs/1991AJ....102..882S} {102, 882}

\bibitem[\protect\citeauthoryear{{Str{\"o}bele} et~al.,}{{Str{\"o}bele}
  et~al.}{2012}]{2012SPIE.8447E..37S}
{Str{\"o}bele} S.,  et~al., 2012, {GALACSI system design and analysis}.
p. 844737, \mn@doi{10.1117/12.926110}

\bibitem[\protect\citeauthoryear{{Takahashi} \& {Inagaki}}{{Takahashi} \&
  {Inagaki}}{1991}]{1991PASJ...43..589T}
{Takahashi} K.,  {Inagaki} S.,  1991, \pasj, \href
  {https://ui.adsabs.harvard.edu/abs/1991PASJ...43..589T} {43, 589}

\bibitem[\protect\citeauthoryear{{Tiongco}, {Vesperini}  \& {Varri}}{{Tiongco}
  et~al.}{2017}]{2017MNRAS.469..683T}
{Tiongco} M.~A.,  {Vesperini} E.,   {Varri} A.~L.,  2017, \mn@doi [\mnras]
  {10.1093/mnras/stx853}, \href
  {https://ui.adsabs.harvard.edu/abs/2017MNRAS.469..683T} {469, 683}

\bibitem[\protect\citeauthoryear{{VandenBerg}, {Brogaard}, {Leaman}  \&
  {Casagrande}}{{VandenBerg} et~al.}{2013}]{2013ApJ...775..134V}
{VandenBerg} D.~A.,  {Brogaard} K.,  {Leaman} R.,   {Casagrande} L.,  2013,
  \mn@doi [\apj] {10.1088/0004-637X/775/2/134}, \href
  {http://adsabs.harvard.edu/abs/2013ApJ...775..134V} {775, 134}

\bibitem[\protect\citeauthoryear{{Vasiliev}}{{Vasiliev}}{2019}]{2019MNRAS.489..623V}
{Vasiliev} E.,  2019, \mn@doi [\mnras] {10.1093/mnras/stz2100}, \href
  {https://ui.adsabs.harvard.edu/abs/2019MNRAS.489..623V} {489, 623}

\bibitem[\protect\citeauthoryear{{Vogt} et~al.,}{{Vogt}
  et~al.}{1994}]{1994SPIE.2198..362V}
{Vogt} S.~S.,  et~al., 1994, {HIRES: the high-resolution echelle spectrometer
  on the Keck 10-m Telescope}.
p.~362, \mn@doi{10.1117/12.176725}

\bibitem[\protect\citeauthoryear{{Watkins}, {van der Marel}, {Bellini}  \&
  {Anderson}}{{Watkins} et~al.}{2015}]{2015ApJ...803...29W}
{Watkins} L.~L.,  {van der Marel} R.~P.,  {Bellini} A.,   {Anderson} J.,  2015,
  \mn@doi [\apj] {10.1088/0004-637X/803/1/29}, \href
  {https://ui.adsabs.harvard.edu/abs/2015ApJ...803...29W} {803, 29}

\bibitem[\protect\citeauthoryear{{Weilbacher} et~al.,}{{Weilbacher}
  et~al.}{2020}]{2020A&A...641A..28W}
{Weilbacher} P.~M.,  et~al., 2020, \mn@doi [\aap]
  {10.1051/0004-6361/202037855}, \href
  {https://ui.adsabs.harvard.edu/abs/2020A&A...641A..28W} {641, A28}

\bibitem[\protect\citeauthoryear{{de Boer}, {Gieles}, {Balbinot},
  {H{\'e}nault-Brunet}, {Sollima}, {Watkins}  \& {Claydon}}{{de Boer}
  et~al.}{2019}]{2019MNRAS.485.4906D}
{de Boer} T.~J.~L.,  {Gieles} M.,  {Balbinot} E.,  {H{\'e}nault-Brunet} V.,
  {Sollima} A.,  {Watkins} L.~L.,   {Claydon} I.,  2019, \mn@doi [\mnras]
  {10.1093/mnras/stz651}, \href
  {https://ui.adsabs.harvard.edu/abs/2019MNRAS.485.4906D} {485, 4906}

\bibitem[\protect\citeauthoryear{{den Brok}, {van de Ven}, {van den Bosch}  \&
  {Watkins}}{{den Brok} et~al.}{2014}]{2014MNRAS.438..487D}
{den Brok} M.,  {van de Ven} G.,  {van den Bosch} R.,   {Watkins} L.,  2014,
  \mn@doi [\mnras] {10.1093/mnras/stt2221}, \href
  {https://ui.adsabs.harvard.edu/abs/2014MNRAS.438..487D} {438, 487}

\bibitem[\protect\citeauthoryear{{van den Bosch}, {de Zeeuw}, {Gebhardt},
  {Noyola}  \& {van de Ven}}{{van den Bosch}
  et~al.}{2006}]{2006ApJ...641..852V}
{van den Bosch} R.,  {de Zeeuw} T.,  {Gebhardt} K.,  {Noyola} E.,   {van de
  Ven} G.,  2006, \mn@doi [\apj] {10.1086/500644}, \href
  {https://ui.adsabs.harvard.edu/abs/2006ApJ...641..852V} {641, 852}

\bibitem[\protect\citeauthoryear{{van den Bosch}, {van de Ven}, {Verolme},
  {Cappellari}  \& {de Zeeuw}}{{van den Bosch}
  et~al.}{2008}]{2008MNRAS.385..647V}
{van den Bosch} R.~C.~E.,  {van de Ven} G.,  {Verolme} E.~K.,  {Cappellari} M.,
    {de Zeeuw} P.~T.,  2008, \mn@doi [\mnras]
  {10.1111/j.1365-2966.2008.12874.x}, \href
  {https://ui.adsabs.harvard.edu/abs/2008MNRAS.385..647V} {385, 647}

\bibitem[\protect\citeauthoryear{van~der Walt, Colbert  \& Varoquaux}{van~der
  Walt et~al.}{2011}]{numpy}
van~der Walt S.,  Colbert S.~C.,   Varoquaux G.,  2011, \mn@doi [Computing in
  Science \& Engineering] {10.1109/MCSE.2011.37}, 13, 22

\makeatother
\end{thebibliography}

\bsp	
\label{lastpage}
\end{document}